\documentclass[aps,pra,twocolumn]{revtex4-1}        
\usepackage{dcolumn}
\usepackage{multirow}
\usepackage{graphicx}
\usepackage{bbold}
\usepackage{subfigure}
\usepackage{epsf}
\usepackage{epstopdf}
\usepackage{color}

\begin{document}
\title{Vibrational stability of graphene under combined shear and axial strains}
\author{Giulio Cocco} 
\author{Vincenzo Fiorentini}
\affiliation{Dipartimento di Fisica dell'Universit\`a di Cagliari and CNR-IOM, Cittadella Universitaria, Monserrato, I-09042 Cagliari, Italy}
\date{\today}
\begin{abstract}
We study the vibrational  properties of graphene under combined shear and uniaxial tensile strain using  density-functional perturbation theory. Shear strain always causes rippling instabilities with strain-dependent direction and wavelength; armchair strain contrasts this instability, enabling  graphene stability in large range of combined strains. A complementary description based on membrane elasticity theory nicely clarifies the competition of shear-induced instability and uniaxial tension.  We also report the large strain-induced shifts of the split components of the $G$ optical phonon line, which may serve as a shear diagnostic.
As to the electronic properties, we find that conical intersections move away from the Brillouin zone border under strain, and tend to coalesce at large strains, making the  opening of gaps difficult to assess. By a detailed search, we find that even at large strains only small gaps in the tens-of-meV range open at the  former Dirac points.
  \end{abstract}
\maketitle

\section{Introduction}
Ideal free-standing and unstrained graphene is, strictly speaking, an abstraction. Real graphene is invariably subjected to strain, either intentional or unintentional, as it gets manipulated, deposited on a substrate, and attached to, or suspended from nanostructured devices.  
Importantly, a general strain may trigger vibrational instabilities, such as long-wavelength rippling, and affect qualitatively the   band dispersion near the Fermi energy. 
Previous work  \cite{tensioni,indiani, cocco} has suggested that uniaxial or biaxial strains below about 25\% do not cause either a gap opening or rippling. While dynamical properties have been studied for axial strains   (see e.g. \cite{tensioni,indiani}),  shear strain has been,  with a few exceptions \cite{zerofield,russi,sheartrans,ripplemd,rippledftb}, largely neglected so far. Only one recent work studies the stability of graphene (modeled with empirical potentials) under combined shear-tensile strains   \cite{russi}  and finds a drastically reduced  stability region in strain space; specifically, the last ``stability island'' at large strain occurs for  a combination of armchair and shear strains.   

In this paper, we consider the effects of shear strain, by itself and in  combination with  armchair uniaxial strain, on the vibrational stability of graphene from a first-principles theoretical perspective, using density-functional perturbation  theory. We find that  shear makes graphene unstable against  a fairly short-wavelength rippling with direction depending on strain intensity, whereas graphene remains stable under a not too large combined strain; the results are consistent with the above-mentioned empirical-interatomic-potentials study  \cite{russi}. We also provide arguments from the elasticity theory of membranes, which confirms the essence of our results. In terms of the electronic properties, we find that only small gaps open up at the Dirac cones, which are displaced away from the  Brillouin zone border.

\section{Method}
 Our ab initio calculations are done within density-functional perturbation theory  \cite{dfpt} 
in the local density approximation using the {\sc Quantum ESPRESSO} code  \cite{qe}. Ultrasoft pseudopotentials  \cite{pp} and a plane-wave basis with principal cutoff 37 Ryd and charge cutoff 450 Ryd  were used. For each strain state we fully relax the internal coordinates in the primitive cell of graphene until all force components are below a stringent threshold of 0.5 meV/\AA. We then perform the phonon calculation on the relaxed structure. Negative squared frequencies (i.e. imaginary frequencies) signal instabilities, providing also their wavelength and spatial pattern. This is the standard procedure in the search for instabilities in high-symmetry phases (see, e.g., Refs.\cite{FE} and  \cite{rond} for the typical case of ferroelectric perovskites).
After extensive testing, we settle on a k-space integration mesh of 16$\times$16$\times$1 for both energy selfconsistency and phonon calculations, with a cold smearing of 0.03 Ryd. Such a rather fine mesh is found to be necessary (and sufficient) to describe the much slower decay of the interatomic force constants in graphene compared to e.g. diamond  \cite{marzari}. Indeed, coarser meshes tend to produce  spurious instabilities in unperturbed graphene. 

\begin{figure}[ht]
\includegraphics[clip,width=7.5cm]{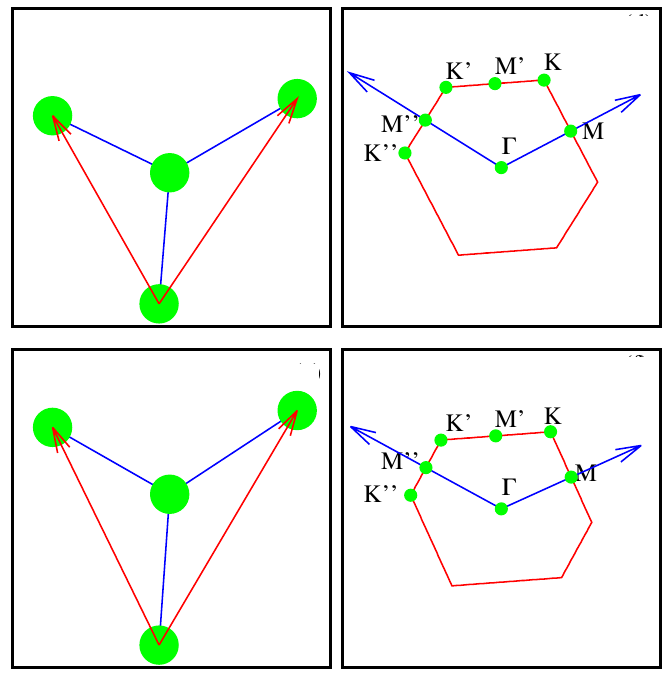}
\caption{(Color online) Primitive cells (left) and Brillouin zones (right) for graphene under shear (top) and combined shear-armchair strain (bottom). The degree of distortion amplified to be visually appreciable.}
\label{fig1}
\end{figure}

Our choice   of the graphene primitive vectors is
\begin{equation}
{\bf a}_1=\frac{1}{2}a\hat{\bf x}+\frac{\sqrt{3}}{\,2}a\hat{\bf y},\ 
{\bf a}_2=-\frac{1}{2}a\hat{\bf x}+\frac{\sqrt{3}}{\,2}a\hat{\bf y},\ 
\end{equation}
with $a$=2.448 \AA.
We apply   a combination of shear and axial strains   multiplying  the primitive vectors by the matrix  
\begin{equation}
S=\mathbb{1}+\left( \begin{array}{cc}
\zeta_{zz} &   \zeta_{sh}\\
 \zeta_{sh} & \zeta_{ac}
\end{array} \right).
\end{equation}
Pure shear is obtained for vanishing armchair and zig-zag strains, i.e. $\zeta_{ac}$=$\zeta_{zz}$=0. A generic point {\bf r}=($x$,$y$) under the action of  $S$ becomes $S${\bf r}={\bf r}+{\bf u}, with 
\begin{equation}
{\bf u}= (\zeta_{sh} y + \zeta_{zz} x, \zeta_{sh} x + \zeta_{ac} y).
\label{ustr}
\end{equation}
We do not apply strain along the  zig-zag direction, hence set $\zeta_{zz}$=0. The shear and armchair strains in standard usage (e.g. in Ref.\onlinecite{russi}) are then given by $\epsilon_{xy}$=2$\zeta_{sh}$  
and $\epsilon_{yy}$=$\zeta_{ac}$. Besides  pure shear, here we consider combined strains in the vicinity of $\epsilon_{xy}$=$\epsilon_{yy}$.
Strain states are labeled by ($\epsilon_{xy}$,$\epsilon_{yy}$) with strains expressed as percentage; for example $\epsilon_{xy}$=0.02, $\epsilon_{yy}$=0.03 is labeled (2,3). 
The primitive cell and Brillouin zone for  graphene under shear and combined strain is shown in Fig.\ref{fig1}.  Shear lowers the symmetry of graphene  to the $C_{2h}$ point group and space group  2/$m$. The six equivalent $K$ points where the Dirac cone is located in pristine graphene become now three pairs of inequivalent points $K$, $K'$, and $K''$ under strain. The lattice vectors both rotate clockwise under  shear (rate 8$'$/\%, i.e. about 1$^{\circ}$ at 7\%);  under combined strains such as ($n$,$n$), {\bf a}$_1$ rotates counterclockwise (rate 5$'$30$''$/\%) and {\bf a}$_2$ clockwise (rate 22$'$16$''$/\%).

\section{Results}

\subsection{Vibrational stability}
\begin{figure}[ht]
\centering
\includegraphics[clip,width=8cm]{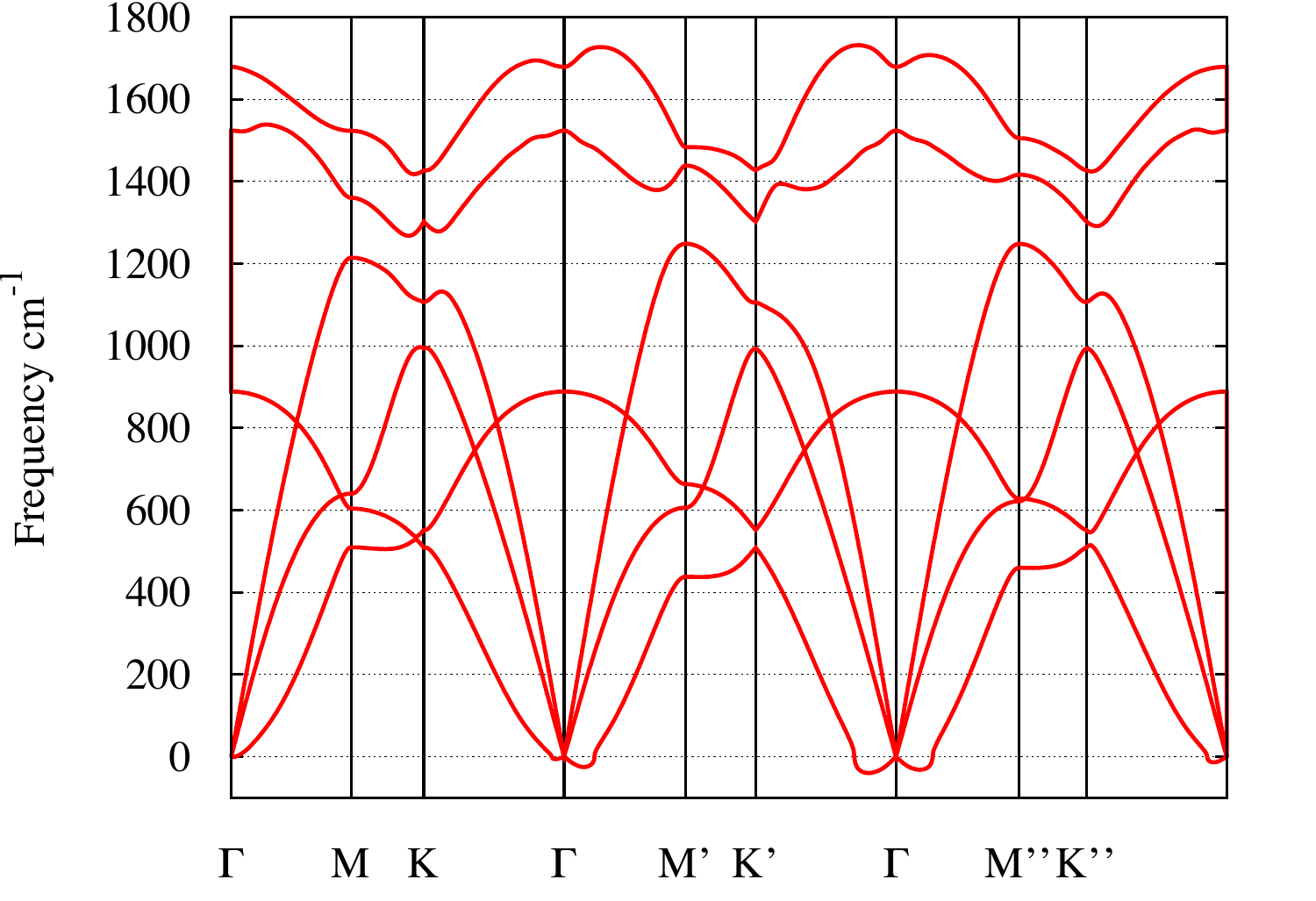}
\caption{(Color online) Phonon dispersion for strain state (3,0), i.e. pure shear at 3\%. Regions of long-wavelength instability are clearly visible. Imaginary frequencies are drawn as negative.}
\label{fig2}
\end{figure}

\begin{figure}[ht]
\centering
\includegraphics[clip,width=4.2cm]{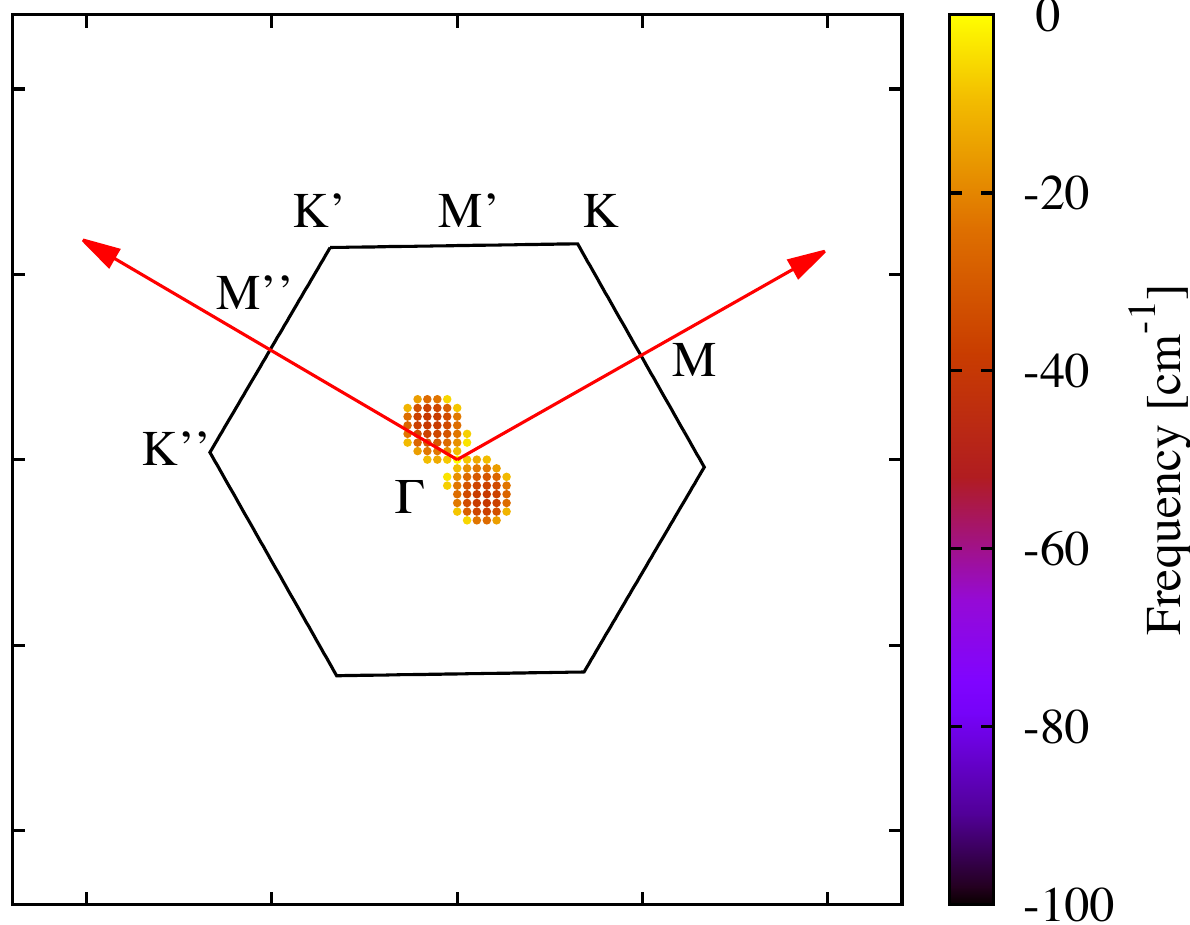}
\includegraphics[clip,width=4.2cm]{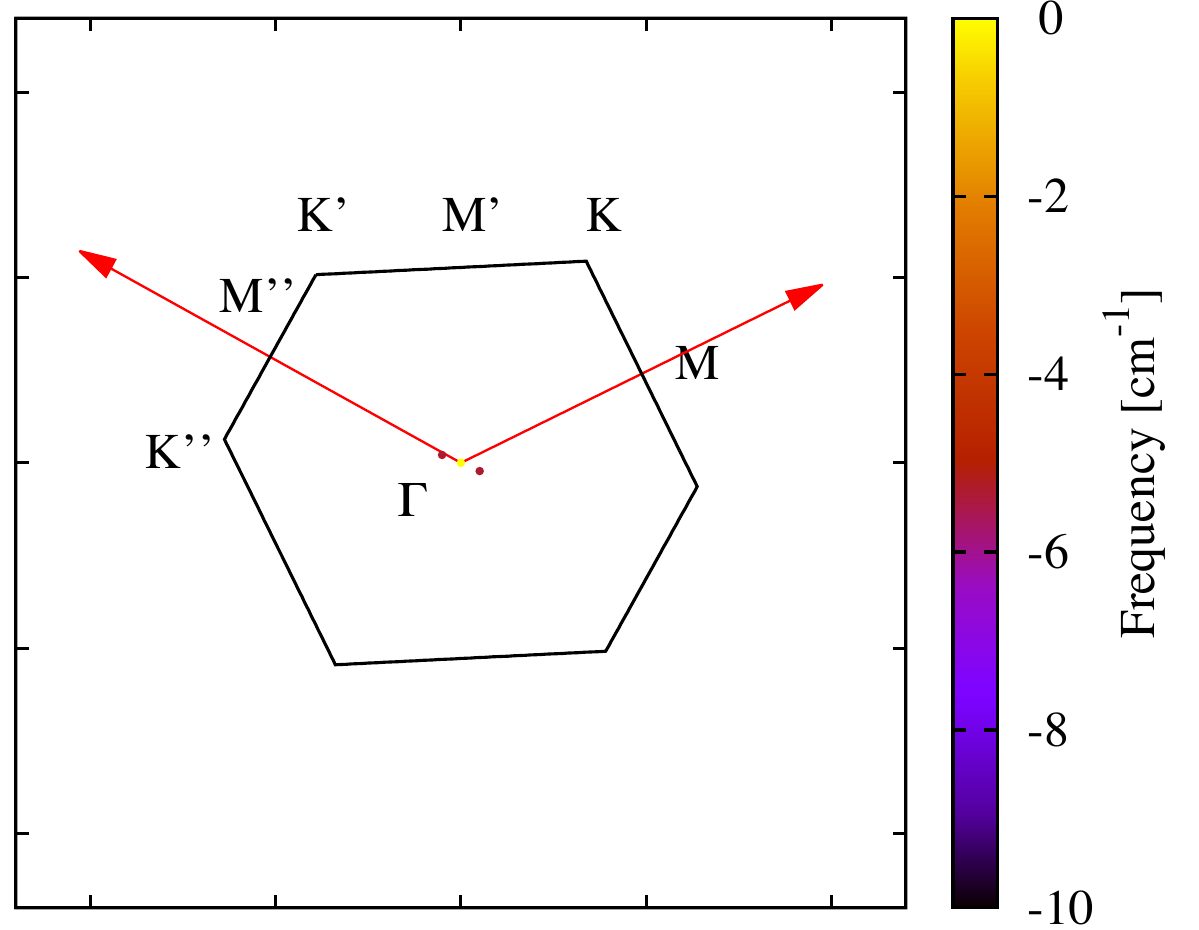}
\caption{(Color online) Contour map of the imaginary frequencies (again conventionally drawn as negative) for strain states (3,0), left panel, and (10,10), right panel. The former has a fully developed instability, while the latter would seem at most only marginally unstable.}
\label{fig3}
\end{figure}

The phonon dispersion for unperturbed graphene (not shown) is in good agreement with previous calculations. Under all the shear strains investigated,  the phonon dispersion exhibits imaginary frequencies near zone center, signaling long-wavelength vibrational instabilities. This can be seen in Fig.\ref{fig2} in the dispersion for shear 3\%, which is quite typical of all strains investigated, as well as, more clearly, in a contour map of the unstable frequencies in Fig.\ref{fig3}, left panel. The wavevector of the  largest (in modulus) unstable frequency directly provides the wavelength and direction of the distortion pattern that will freeze-in into the graphene lattice upon condensation of the unstable modes. It is a rippling pattern with wavelength decreasing with shear, namely 17.3, 15.4, 12.6, 10.6 \AA\, for shear 3, 5, 7, 10 \%; the normal to the rippling wavefront backs away  counterclockwise from the shear axis (the bisector of the direct lattice vectors) for increasing strain, towards an angle of 45$^{\circ}$ from the axis. 

\begin{figure}[ht]
\centering
\includegraphics[clip,width=8cm]{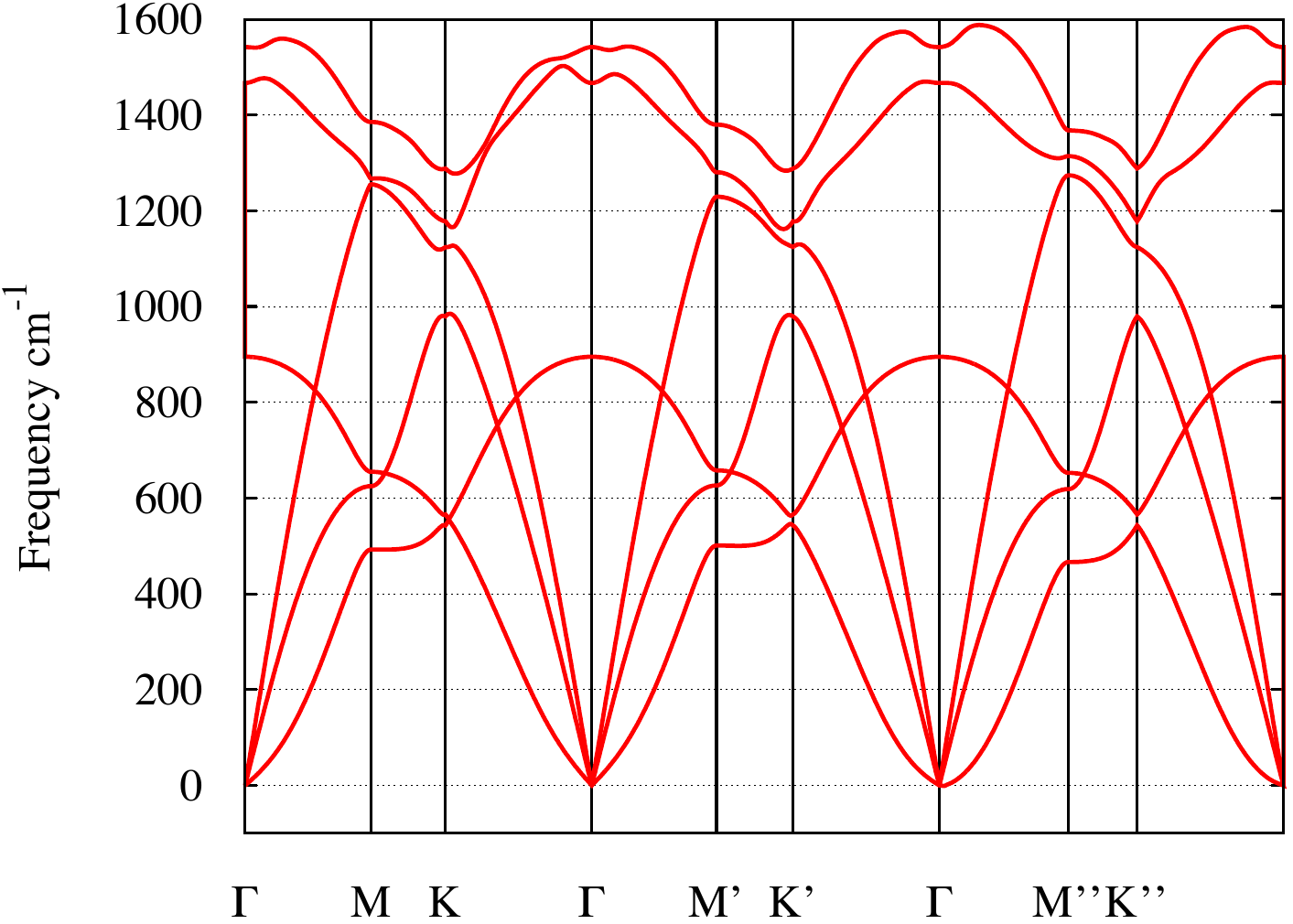}
\caption{(Color online) Phonon dispersion for combined-strain state (3,3). No instability  is present.}
\label{fig4}
\end{figure}

For combined shear and armchair-tensile strain, at strains below about 10\% there appears to be no instability. The quartic dispersion $\omega^2$$\simeq$$q^4$ of the flexural ZA mode  \cite{lif,lif2,fas,fasolino} acquires a quadratic term that increases with strain amplitude. As can be seen for the typical case of the (3,3) strain in Fig.\ref{fig4}, and not unexpectedly, the dispersion is significantly anisotropic. At 10\% combined strain (Fig.\ref{fig3}, right panel) a marginal instability appears; its phase space is  very limited  and  its condensation will lead to a small energy gain. The wavelength in this case is 45-50 \AA, and the angle is much larger than for pure shear, i.e. 160$^{\circ}$ from the horizontal zig-zag chain direction (Fig.\ref{fig1}) at zero strain. 

\begin{figure}[ht]
\centering
\includegraphics[clip,width=8cm]{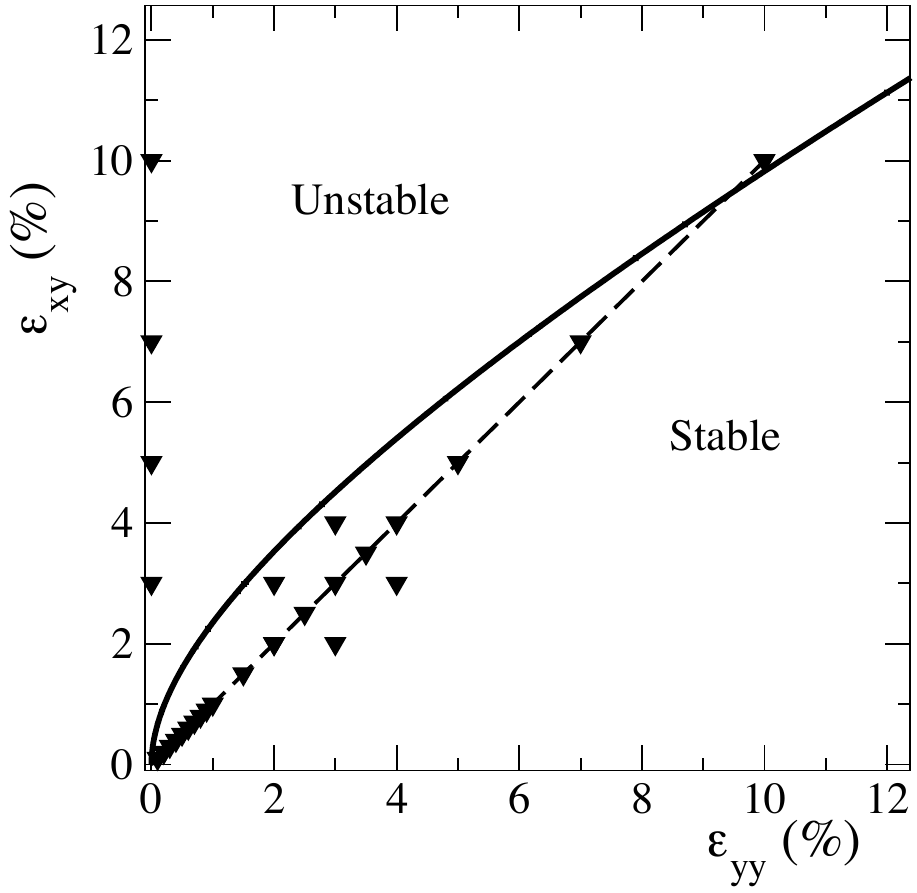}
\caption{Schematic phase diagram for vibrational stability. Triangles are calculated strains.}
\label{fig9}
\end{figure}

Previous searches for rippling by ab initio methods under uniaxial strain produced negative results  \cite{indiani} (the reason will become clear in Sec.\ref{elast}). On the other hand, the only work on combined strains  \cite{russi} using empirical potentials suggests  (Fig.3 of Ref.\onlinecite{russi}) that shear strain reduces the graphene stability region in strain space. Our results are in general agreement with this conclusion; not only does instability occur at any pure shear strain, but the strain of 10\%, where we barely glimpse an instability, is indeed at the border of stability as predicted in Ref.\onlinecite{russi}. From our calculations we infer the qualitative phase diagram for stability drawn in Fig.\ref{fig9}, which in this region is similar to the relevant section of Fig.3 of Ref.\onlinecite{russi}.
 The qualitative conclusion is that shear destabilizes graphene, whereas if uniaxial tension is blended in,  stability is re-established in a vast region of strain space. We discuss this further in the next Section.
(We recall that vibrational  instability should abruptly set in at a pure tension $\epsilon_{yy}$$\sim$20\%  according to empirical as well as ab initio calculations  \cite{russi,indiani}.)

\begin{figure}[ht]
\centering
\includegraphics[clip,width=7cm]{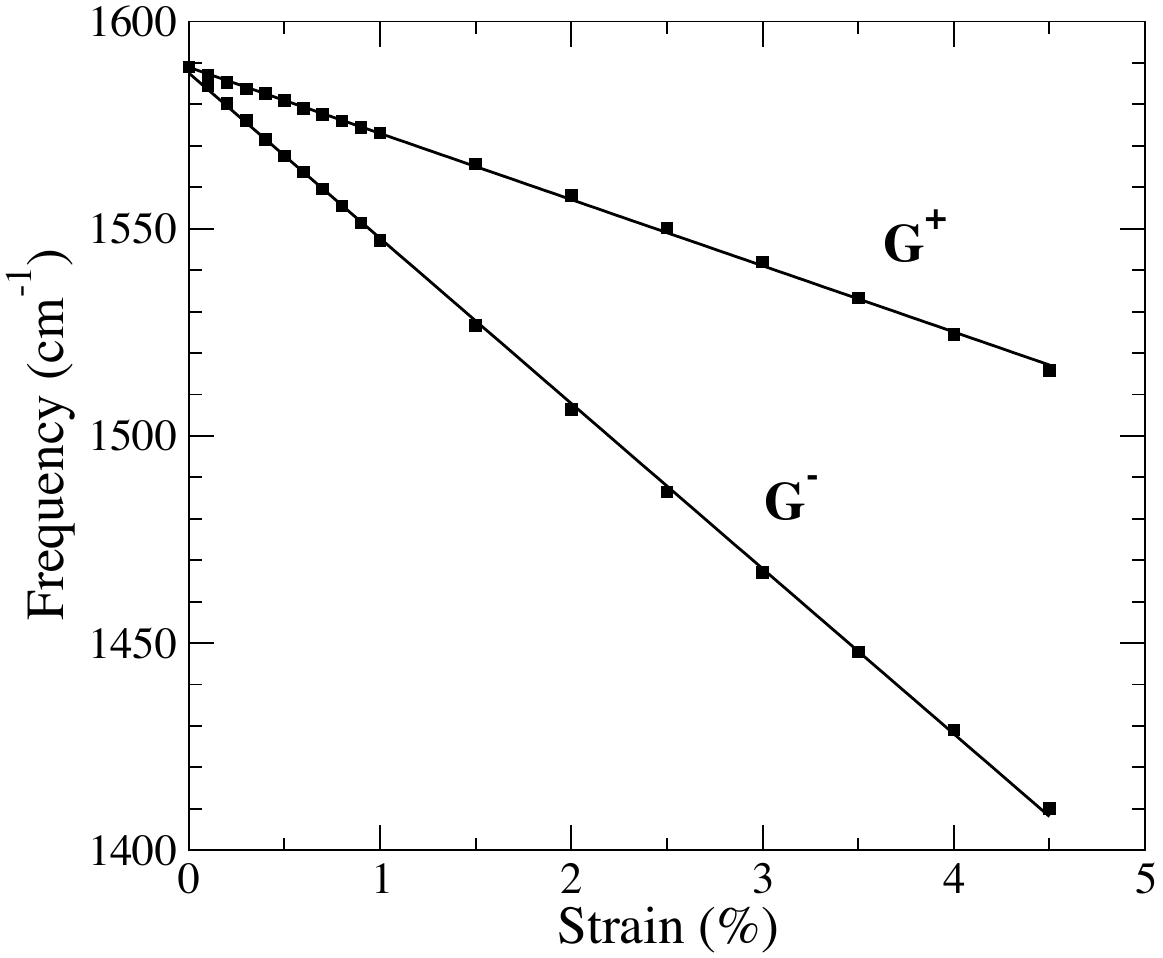}
\caption{Position of strain-split optical-phonon lines vs combined strain.}
\label{fig5}
\end{figure}

An interesting byproduct  of our combined-strain calculations is the strain derivative of the energies of the optical ``$G$'' lines resulting from the strain splitting of the optical phonon. For uniaxial strain, these slopes were measured  \cite{ggp} by Raman scattering to be --10.8 and --31.7 cm$^{-1}$/\%. As shown in Fig.\ref{fig5}, obtained by additional calculations at small combined strains, both lines shift linearly. The strain derivatives we obtain are 
--16.0 cm$^{-1}$/\% for the $G^+$ line and --39.8 cm$^{-1}$/\% for the 
$G^-$ line. These values are significantly larger than for uniaxial strains, and could be useful as a diagnostic indicator of the presence of shear in uniaxially strained samples. 

\subsection{Elasticity-based description}
\label{elast}

The analysis of vibrational frequencies of flexural modes based on the elasticity theory of membranes helps rationalize the 
 findings just reported. The elastic energy of a membrane with  bending rigidity $\kappa$, and  Lam\'e coefficients $\mu$ and $\lambda$ is \cite{fasolino}
\begin{equation}
E= \int dA \left[
\frac{\kappa}{2}(\nabla^2 h)^2 +\mu \overline{u}_{\alpha\beta}^2 +\frac{\lambda}{2}\overline{u}_{\alpha\alpha}^2,
\right]
\end{equation}
where 
\begin{equation}
\overline{u}_{\alpha\beta}=\frac{1}{2}\left(\partial_{\beta}u_{\alpha}+
\partial_{\alpha}u_{\beta} + \partial_{\alpha}h\partial_{\beta}h
\right)
\end{equation}
is the strain tensor, $h$ is the vertical displacement, and the vector {\bf u} was defined in Eq.\ref{ustr}. Using a plane wave ansatz $h$=$h_0$\,exp($ik_x x+ik_y y$) for the vertical displacement,  the squared vibrational frequency of the out-of-plane vibration (the flexural modes) is
\begin{equation}
\rho\omega^2 = \kappa k^4 +  (\lambda k^2 + 2\mu k_y^2)\, \epsilon_{yy} + 4\mu k_x k_y\, \epsilon_{xy},
\label{o2}
\end{equation}
where  $\rho$ is the areal mass density, and we have  set $\epsilon_{xx}$=0 as in all previous calculations \cite{nota}.

\begin{figure}[ht]
\centering
\includegraphics[clip,width=4.4cm]{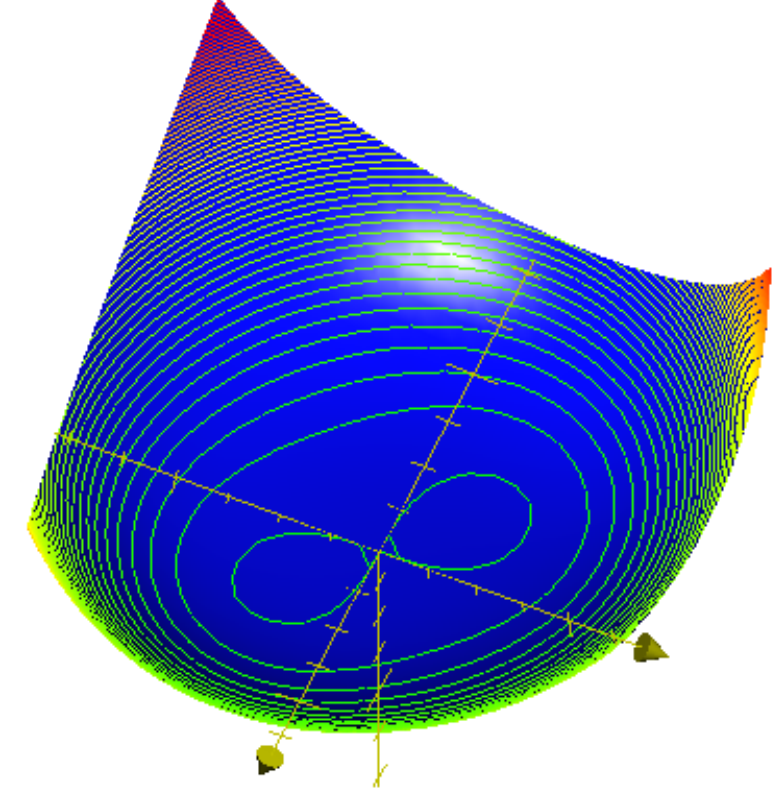}\includegraphics[clip,width=4.4cm]{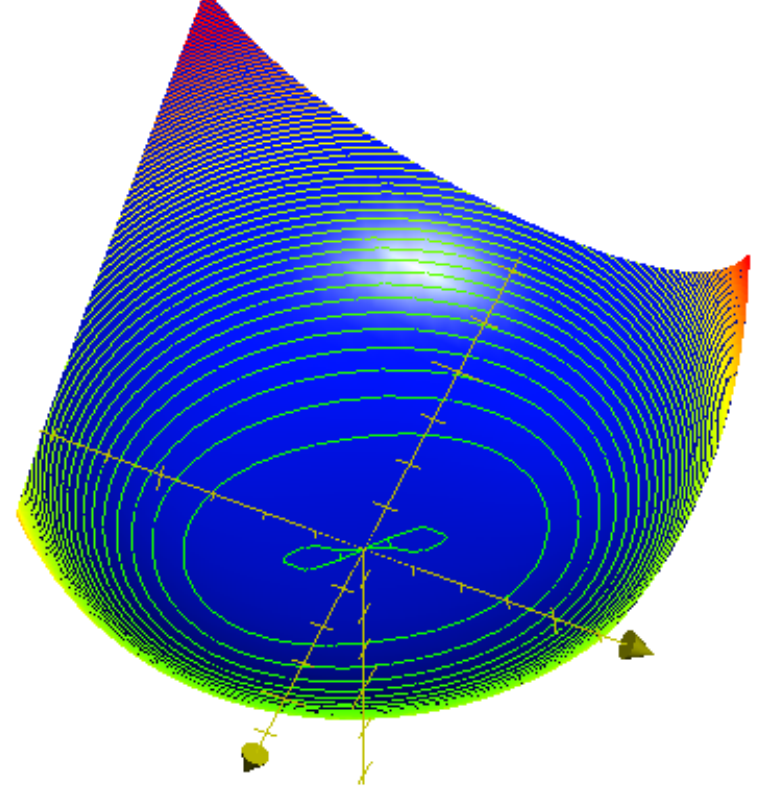}
\caption{(Color online) Sketch of the squared frequency  obtained from elasticity theory for pure shear (left) and nearly compensating shear and uniaxial tension (right). Compare this  view from negative-squared frequencies with the top view in  Fig.\ref{fig3}. In both panels, $\rho$=1, $\kappa$=1.1 eV, $\lambda$=2.57 eV/\AA$^2$, $\mu$=2.4, $\epsilon_{xy}$=0.02. In the left panel, $\epsilon_{yy}$=0; in the right panel, $\epsilon_{yy}$=0.015.}
\label{elastfig}
\end{figure}

In the absence of strain the flexural mode has the expected quartic dispersion. Strain produces quadratic terms that, at small $k$, can prevail over the quartic term and  subvert the normal state of affairs.  The contribution of armchair strain (the second term) is positive and produces no instability, but causes the squared-frequency dispersion to  acquire an anisotropic quadratic term. A pure shear contribution (third term in Eq.\ref{o2})  produces an instability,  because it becomes negative in the second and fourth quadrant of the ($k_x$,$k_y$) plane.  If both  shear and armchair strains are present,  the shear-induced instability is countered by the tensile quadratic term. All these features are indeed seen in the ab initio calculations discussed above. By way of example,  Fig.\ref{elastfig} reports the squared frequency dispersion with pure shear (left) and with nearly compensating shear and  tensile strain (right); these are qualitatively similar to Fig. \ref{fig3}.

Elasticity clearly rationalizes the shear destabilization of graphene, the countervailing action of tension, and the appearance of quadratic terms in the squared-frequency dispersion. To compare the stability range predicted by atomistics and elasticity, we set up a   phase diagram akin to Fig.\ref{fig9} using the elastic moduli  from experiment or simulation, or adjusting them ad hoc (recalculating them is beyond our present scope). The  boundary between stable and unstable regions in our two-dimensional strain space ($\epsilon_{yy}$,$\epsilon_{xy}$), is obtained  counting the negative squared frequencies in $k$ space for each point in strain space, and calculating the ratio R of negative-frequency $k$ points to the total number of points (the sampled $k$ region is a suitable small-$k$ region around $k$=0); any region of strain space where R is non-zero is an instability region, with the zero-R contour line acting as phase boundary. 

\begin{figure}[ht]
\centering
\includegraphics[clip,width=8.5cm]{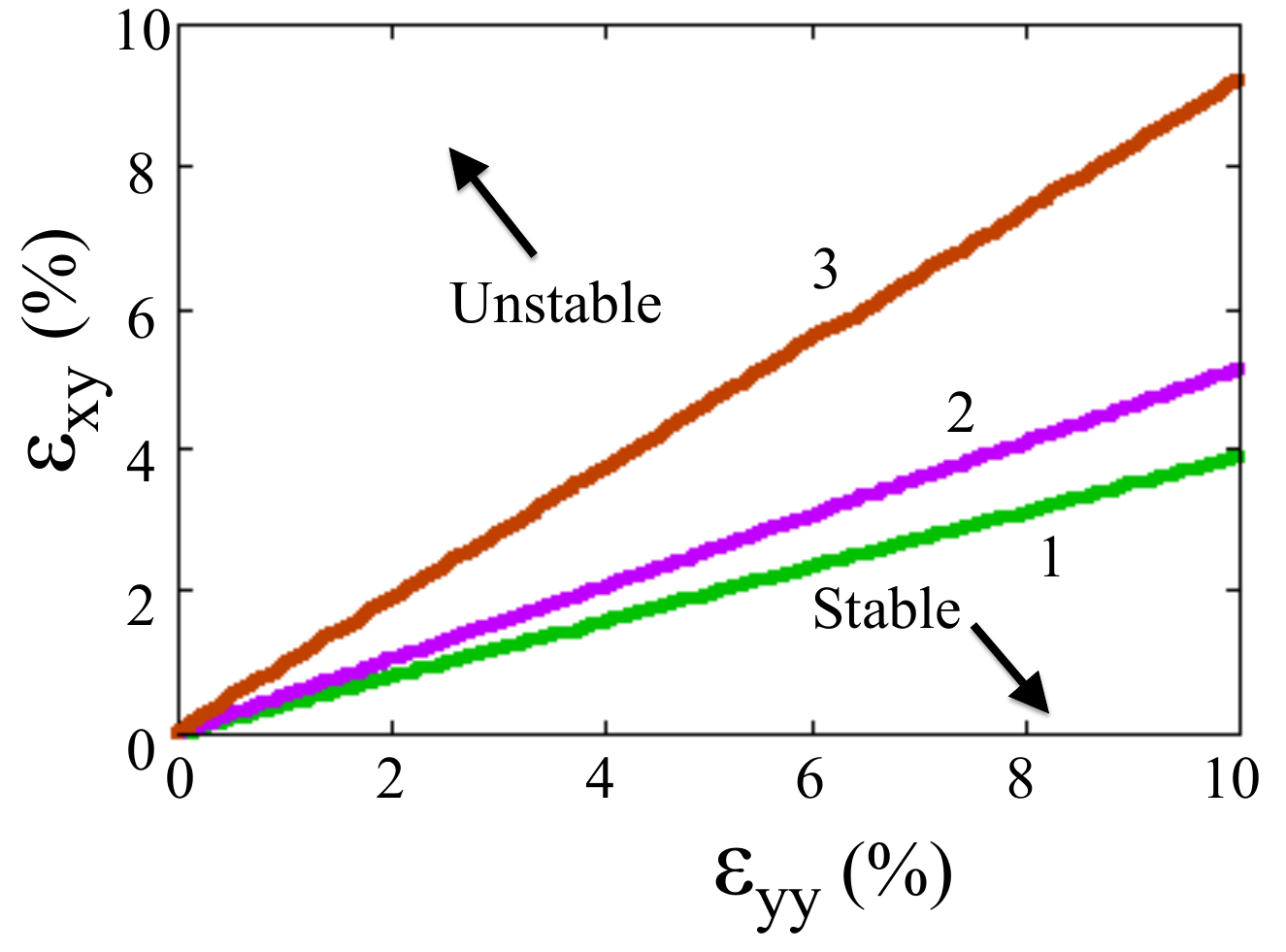}
\caption{(Color online) Phase diagram in strain space from elasticity theory using $\kappa$=1.1 eV,  $\lambda$=2.57 eV/\AA$^2$, and (from bottom to top, labeled 1, 2, 3 respectively) $\mu$=9.95, 6.2, and 2.4 eV/\AA$^2$.}
\label{phb}
\end{figure}

Fig.\ref{phb} shows the phase boundary between vibrationally unstable and stable regions obtained from Eq.\ref{o2} for three combinations of elastic moduli. We set in all three cases $\kappa$=1.1 eV and $\lambda$=2.57 eV/\AA$^2$ \cite{zakh}, but $\mu$ takes on the values (in eV/\AA$^2$) $\mu_1$=9.95   from simulation \cite{zakh}, $\mu_2$=6.2 from the Poisson ratio $\nu$=$\lambda$/($\lambda$+2$\mu$)=0.17 of graphene \cite{book}, and $\mu_3$=2.4 chosen to obtain a phase boundary similar to the atomistic one (Fig.\ref{fig9} and Ref.\cite{russi}). Apparently, for accepted values of the Lam\'e coefficients, elasticity tends to predict a smaller stability region than the ab initio one (which largely agrees with that of semiempirical tight-binding). This may have to do with deviations of the  behavior of microscopically textured graphene from that of an elastic continuum (as suggested  earlier  \cite{fasolino} in relation to long-wavelength thermal fluctuations  \cite{thfl}). Aside from these moderate quantitative deviations, however, elasticity and ab initio results largely agree and  provide useful, mutually complementary information. 

\subsection{Band structure}

For all the strains containing a shear component, the appearance of a gap at the conical intersections at the vertexes of the Brillouin zone is guaranteed by symmetry. Shear strain lowers the symmetry of graphene to the space group 2/$m$; hence the little group of any $k$ point in the Brillouin zone can be at most the point group $C_{2h}$,  whose  irreducible representations are all one-dimensional. Therefore, band degeneracies are not protected by symmetry anywhere in the Brillouin zone; the gap-opening region in the phase diagram is all of Fig.\ref{fig9} or \ref{phb} (except the $\epsilon_{yy}$ axis where shear vanishes). However, it is unknown  a priori how large  the gap will be  under a specific strain intensity.

We examined the band structure in a number of strain states using an uncommonly  dense $k$-point sampling to pinpoint the displacement of the conical intersections, and the nature and value of the gaps.
In essence, for the strains of interest here, we find that:  
a) the conical intersections move away from the corners and, in fact, even from the border of the strain-distorted zone; this can lead to qualitative errors if only the usual path around the Brillouin zone is explored  \cite{error-bands};  b) at large strains the Brillouin zone distorts toward a rhomboidal shape, and accordingly the former Dirac points become nearer and nearer, and tend to coalesce at the largest strains; c) despite the loss of hexagonal symmetry, the conical band structure does largely survive as such up to very large strains. Indeed,  a gap does exist, but it is always small, and accordingly the  parabolic section at the band minimum is  very localized  in k space. 

\begin{figure}[ht]
\centering
\includegraphics[clip,width=4cm]{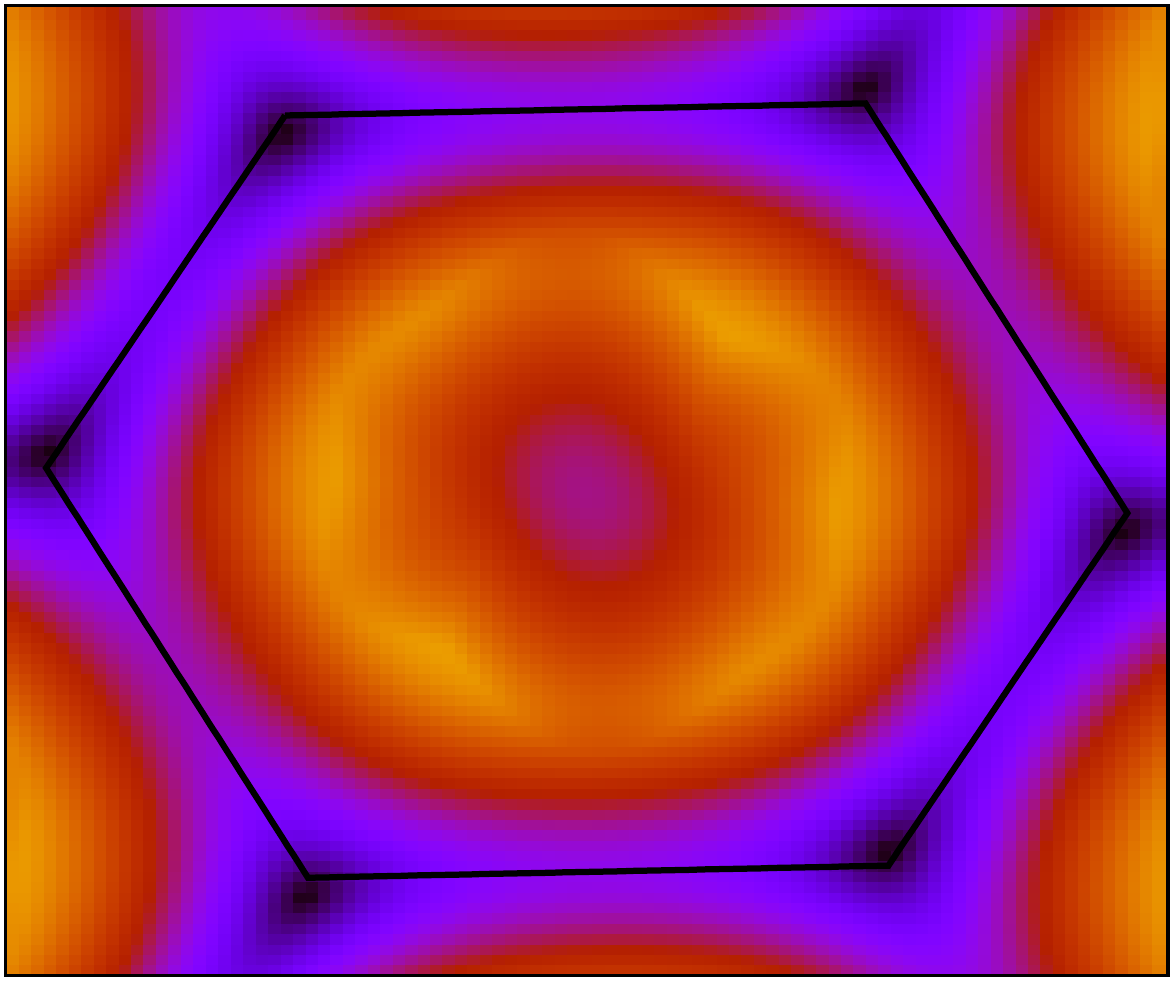}
\includegraphics[clip,width=0.405cm]{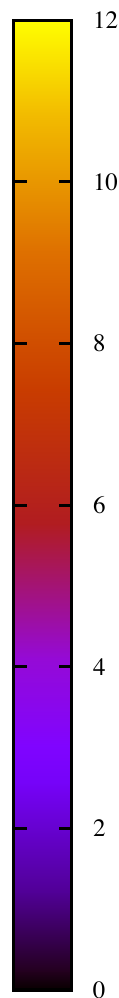}
\includegraphics[clip,width=4cm]{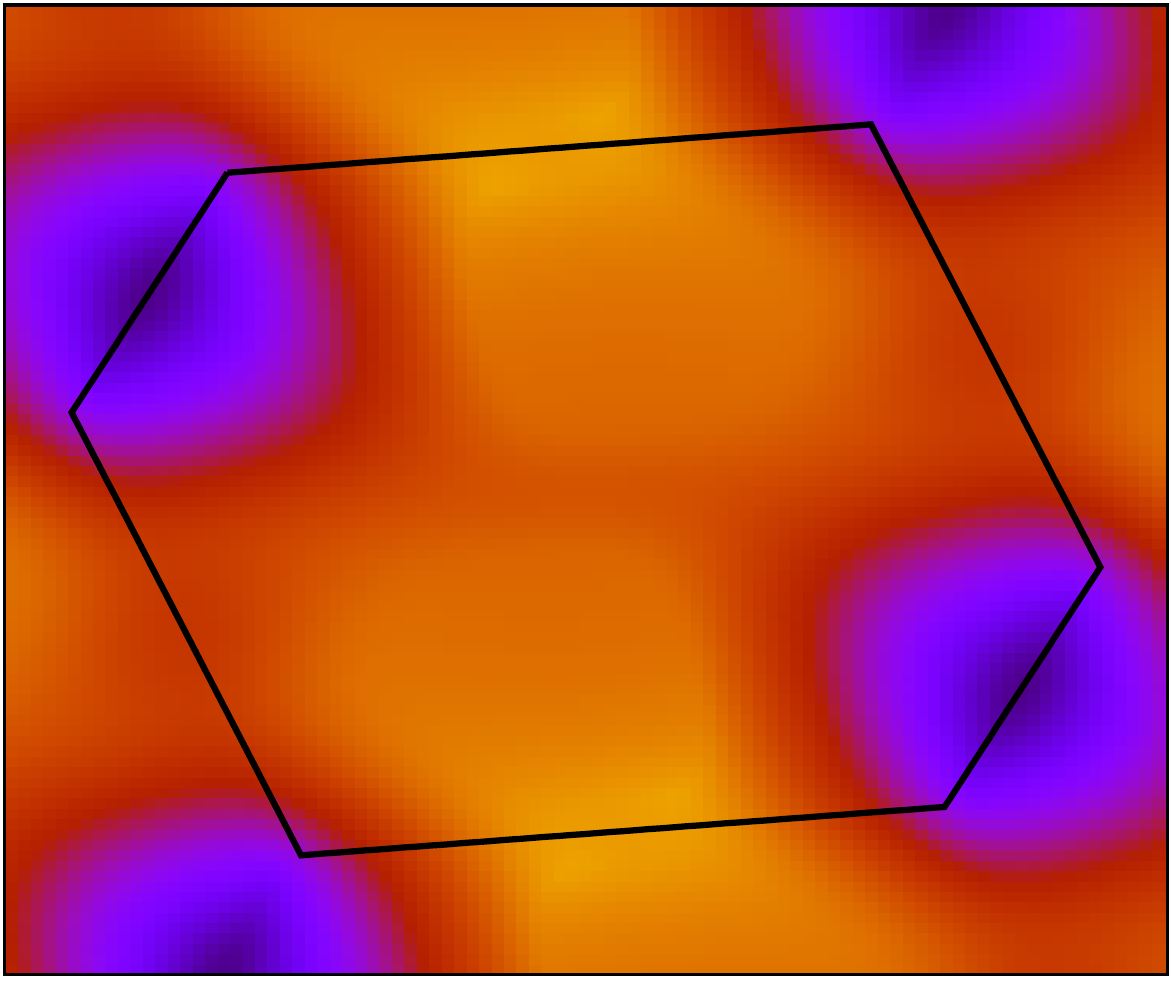}\\
\includegraphics[clip,width=4cm]{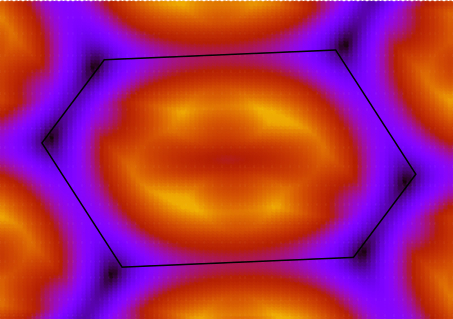}
\includegraphics[clip,width=0.34cm]{./fig9e.pdf}
\includegraphics[clip,width=4cm]{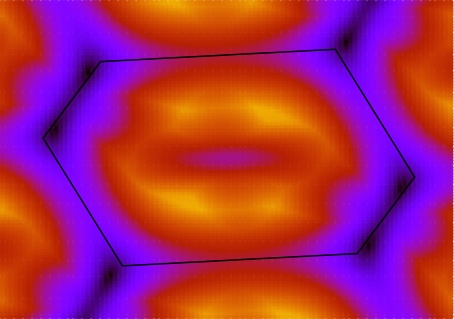}

\caption{(Color online)  Gap value near the former Dirac points in the  Brillouin zone under increasing combined  shear-tensile strain. Counterclockwise from top left:  (5,5), (12,18), (15,20), (18,18). The Dirac points migrate away from zone borders and tend to coalesce at large combined strains. The color-code scale unit is 10 meV.}
\label{fig7}
\end{figure}

As to points a) and b), in Fig.\ref{fig7} we show the gap (i.e. the eigenvalue difference of the last occupied and first empty band) vs wavevector as a contour map in a large portion of k-space. Four strains are displayed in Fig.\ref{fig7}, counterclockwise from top left:  (5,5), (12,18), (15,20), and (18,18). These four strains states are within the stability region according to Ref.\onlinecite{russi}; the last one may fall outside our own stability region (Fig.\ref{fig9}), although extrapolation is of course  quite uncertain.
As this strain series illustrates, the conical intersections, identified by the small gap,  are displaced away from the corners and border of the zone and eventually, at large strains, tend to coalesce.   
  
\begin{figure}[ht]
\centering
\includegraphics[clip,width=7.5cm]{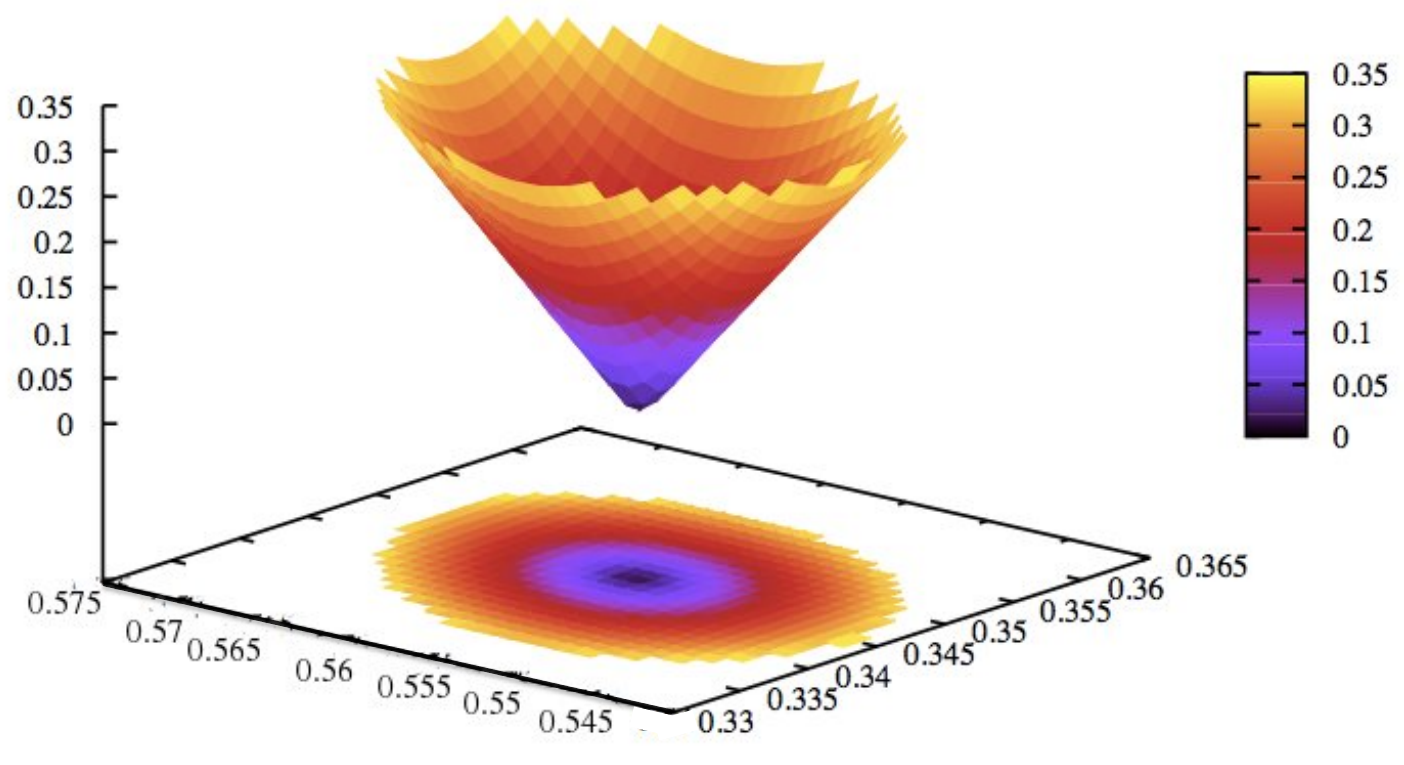}
\caption{(Color online)  Blow-up of the gap near one of the former Dirac points at strain (9,9). The parabolic behavior is visible only at extreme magnification in a region of linear size a few thousandths of the zone. The units are 2$\pi$/$a$ for k and eV for the gap.}
\label{fig8}
\end{figure}

For all the strains we looked at, the gap is  present as symmetry dictates, but never exceeds 0.03-0.04 eV (point c) above).  Pinpointing this behavior requires an extreme level of resolution in $k$-space. Fig.\ref{fig8} shows that a local curvature at the minimum appears near the Dirac point  over a $k$ range of a few thousandths of the linear size of the Brillouin zone, even at the large strain (9,9). The linear portion of the gap function near the Dirac point extrapolates to zero, which  qualifies the  gap as a conventional ``massive" gap rather than a ``massless" gap  \cite{mless}. 
We mention that very recent measurements of the transport properties under shear  \cite{sheartrans}, although somewhat inconclusive as to the detailed effects  of strain,  suggest a metallic character compatible (given that experiment are done at room temperature) with the small gaps just discussed.

The discrepancy with previous \cite{cocco} tight-binding results   (indicating sizable gaps at former Dirac points) may be attributed to the effects of atomic relaxation, which we find to be very substantial especially at large strains. Indeed, during review we  became aware of an unpublished paper  \cite{rippledftb} where rippling occurs when a large supercell under shear and tensile strain is relaxed 
via a semi-empirical tight-binding technique. The shear used is twice as large as the tensile component, and in this strain region we indeed find instability from phonon calculations, as discussed above. The same paper reports a vanishing  gap even under substantial strains. This agrees with our evidence for an (almost) entire suppression of the gap opening due to relaxation. As we have shown, and as dictated by symmetry, the gap is non-zero for combined strains, but it is very small; the use of the density of states in Ref.\onlinecite{rippledftb} may have prevented pinpointing it. 


\begin{figure}[ht]
\centering
\includegraphics[clip,width=8cm]{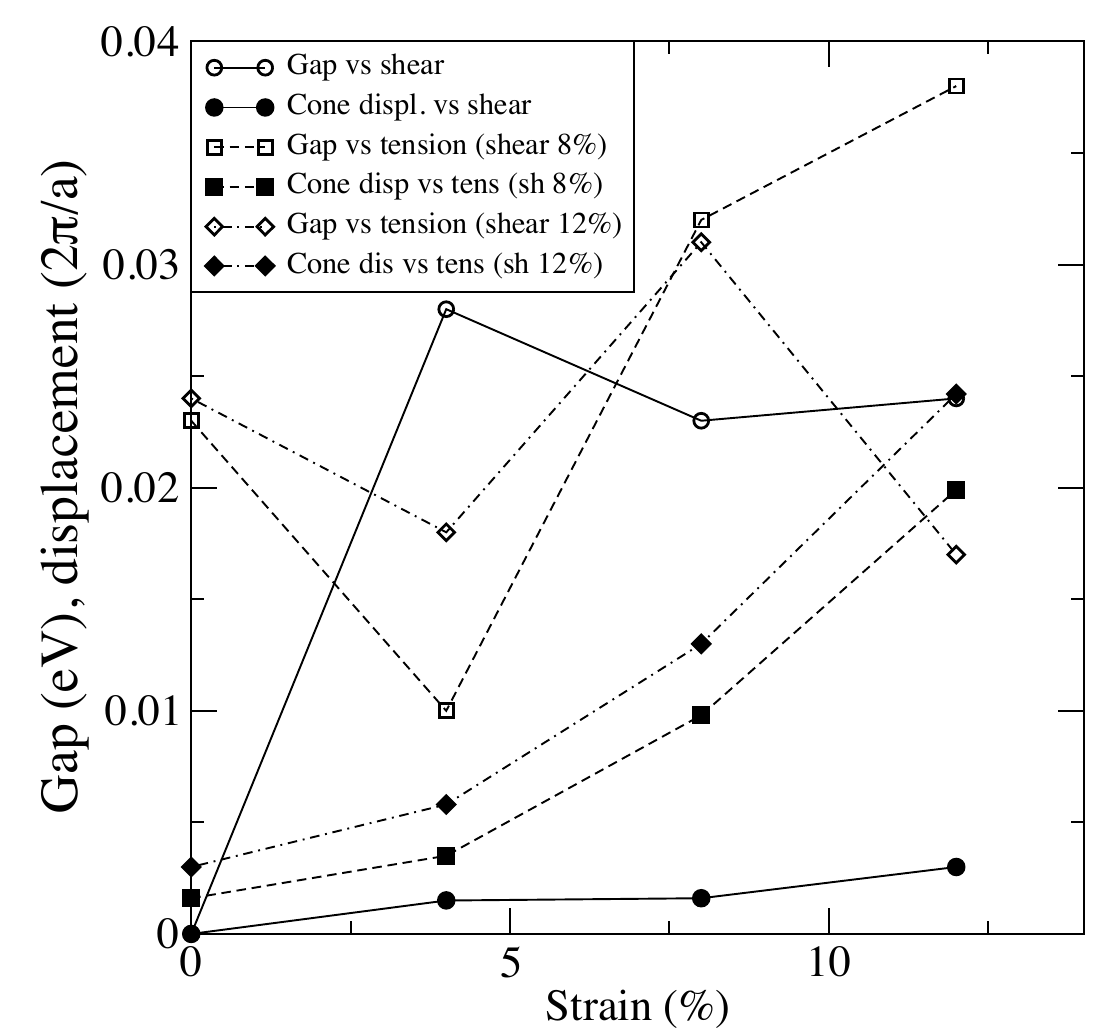}
\caption{(Color online)  Gap (empty symbols) and Dirac-point displacement $\delta$ (filled symbols) vs strain. Circles: gap and $\delta$ vs shear at zero tension; squares: same vs tension at shear 8\%; diamonds: same  vs tension at shear 12\%.}
\label{ultima}
\end{figure}

We close with a  sampling of  the dependence of the gap and its position in k-space on strain intensity. In Fig.\ref{ultima} we show the gap and Dirac point displacement vs shear at zero tension, and vs tension at fixed shear.  The displacement, quantified by the Euclidean distance $\delta$ in k space between the k-point where the gap occurs and the nearest zone corner, clearly increases with strains, more so under combined strains, and reaches several percent even at  moderate strain. The gap, on the other hand,  does not seem to  increase monotonically with any of these strain combination. It is interesting to note that for 4\% shear, at 8\% or 12\% tension, graphene is stable and the gap is 0.04 eV,  larger than that caused by epitaxy on BN \cite{giov}.

\section{Summary}

In summary, ab initio phonon calculations indicate that graphene is stable against rippling instabilities for a wide range of combined armchair-shear strains, whereas pure shear strains always cause instability. Our stability diagram agrees with previous estimates based on empirical potentials.  We  reported the strain derivatives of the strain-split components of the $G$ line, which turn out to be are 25 to 45\% larger than under uniaxial strain, and could be useful as diagnostic measure of  shear strain.
The elasticity theory of membranes helps rationalize the   shear-induced instability and the opposing effect of tension found in ab initio calculations, despite minor quantitative discrepancies.

The electronic gaps expected from symmetry arguments near the Dirac points are always very small at all strains (of order 0.01-0.02 eV). The conical intersections move away from the border of the zone (a potential pitfall in analyzing the band structure), and at large strains they tend to coalesce. A linear band dispersion is largely preserved in the $k$-space vicinity of the former conical intersections. Despite its smallnes, the gap opened by strain may  be of interest to tune the Berry phase (i.e. chirality and quantum Hall phase shift) in typical doping regimes, as we will discuss elsewhere \cite{urru}. By comparison, typical gaps in simple epitaxial graphene systems (for example  \cite{giov} graphene on BN) are also small (30-50 meV), whereas gaps in the 0.1 eV range require complex patterning techniques  \cite{pacile}.

Work supported in part by the MIUR-PRIN 2010 project {\it Oxide}, CAR of Cagliari University, Fondazione Banco di Sardegna and CINECA grants. We are grateful to F. Guinea for motivating the inclusion of the elasticity treatment and for pointing out corrections to the frequency expression.

\end{document}